\documentclass[titlepage]{article}
\begin{document}

\vspace{20cm}
\title{Dark Matter in the\\Local Galactic Disk}
\author{Kerry Capelle Zekser}
\date{Galactic Structure and Stellar Dynamics\\Johns Hopkins University\\May 8, 2002}
\maketitle
\pagebreak

\pagestyle{plain}
\normalsize
\tableofcontents
\pagebreak

\vspace{100mm}

\section{Abstract} \label{sec:abstract}

	The existence of dark matter in the Galactic disk has been a topic of significant debate, especially over the past two decades.  Evidence for dark matter in the disk, a component in addition to the now well-proven dark matter halo, has been meticulously explored by analyzing the local disk density in the solar neighborhood.
  
	Due to the extensiveness of this particular subfield of astrophysics, the goal of this investigation is to gain a broad understanding of the initial debates as well as general insight into some of the more recent advances.  The analysis of the most prominent participants in the debate includes comparing and contrasting the methods, data, assumptions, as well as the ultimate conclusions of each study.  In particular, this discussion introduces the significant and varied issues, such as systematic errors, uncertainties, and selection effects in the data populations, relating to each method of investigating the density and local structure in the solar neighborhood of the Galactic disk.  Furthermore, recent studies involved in the determination of the ultimate conclusion to this quandary are also explored.

\vspace{8mm}
\section{Introduction} \label{sec:intro}

	Dark matter is most often determined from the discrepancy between the visible matter and the total amount of gravitating matter in a system.  Using this approach, the first step in the investigation into the existence of dark matter in the Galactic disk is to determine the total mass density in the local solar neighborhood of the disk; the local volume density, $\rho$, or column density, $\Sigma$.  As discussed in Binney and Tremaine (1994, hereafter BT94), the following is a method of calculating the local matter density which clearly reveals the difficulties in exploring the existence of Galactic disk dark matter.

	In order to derive an expression for the mass density, the Jeans equation, analogous to Euler's equation describing the flow of a fluid, can be derived from the first moment of the collisionless Boltzmann equation with respect to velocity.
\begin{equation} \label{eq:GD4-27}
\nu\frac{\partial\overline{v_j}}{\partial t}+\nu\overline{v_i}\frac{\partial\overline{v_j}}{\partial x_i}+\frac{\partial(\nu\sigma^2_{ij})}{\partial x_i}+\nu\frac{\partial\Phi}{\partial x_j}=0
\end{equation}
When equation \ref{eq:GD4-27} is applied to an axisymmetric system with cylindrical coordinates, this Jeans Equation transforms as follows.  
\begin{equation} \label{eq:GD4-29cmod}
\frac{\partial(\nu\overline{v_z})}{\partial t}+ \frac{1}{R}\frac{\partial}{\partial R}(R\nu\sigma^2_{Rz})+\frac{\partial(\nu\overline{v^2_z})}{\partial z}+\nu\frac{\partial\Phi}{\partial z}=0
\end{equation}
Assuming the Galactic disk is in a steady state, the first term in equation \ref{eq:GD4-29cmod} can be omitted. 
 In addition, as derived in the Appendix Section \ref{sec:app}, the second term in equation \ref{eq:GD4-29cmod} involving the $v_R-v_z$ coupling can also be omitted reducing the problem to one dimension.
\begin{equation} \label{eq:GD4-36}
\frac{1}{\nu}\frac{\partial(\nu\overline{v^2_z})}{\partial z}=-\frac{\partial\Phi}{\partial z}
\end{equation}
The omission of the term relating the $v_R-v_z$ coupling was addressed by Kuijken and Gilmore (1989a, hereafter KG89a) in which a final correction for this effect was made; this correction is discussed in Section \ref{sec:con2}.  

Because Poisson's Equation for a thin disk simplifies to 
\begin{equation} \label{eq:GD4-37}
\frac{\partial^2\Phi}{\partial z^2}=4\pi G\rho
\end{equation}
We can transform equations \ref{eq:GD4-36} and \ref{eq:GD4-37} into one equation independent of the potential:
\begin{equation} \label{eq:GD4-38mod}
\rho=-\frac{1}{4\pi G}\frac{\partial}{\partial z}\bigg[\frac{1}{\nu}\frac{\partial(\nu\overline{v^2_z})}{\partial z}\bigg]
\end{equation}
Therefore, the mass density can be estimated for a particular stellar population once determination is made of the number density, $\nu$, and the mean-square vertical velocity, $\overline{v_z^{2}}$, as a function of height from the Galactic plane.  This was the method originally utilized by Oort to estimate the local Galactic disk density, known now as the Oort limit, $\rho_0=0.15 M_\odot /pc^3$.  By exploring the estimated mass contribution from stars, remnants, stellar gas, and dust, an approximation of the local observed mass density by BT94 was found to be on the order of $\rho_0=0.114 M_\odot /pc^3$.  The discrepancy indicates the existence of dark matter which is not accounted for observationally.

	The main problem with the previous method is that this estimation is highly uncertain due to the three derivatives involved in this determination of the mass density.  Firstly, the number density is calculated by differentiating the stellar counts of a particular population tracer such as K giants or K dwarfs; in particular, the differentiation of data involving error bars contains considerable uncertainty.  In addition, there are the other two obvious differentiations in equation \ref{eq:GD4-38mod}.  

	Alternatively, the column density within a couple scale heights of the Galactic plane is often investigated to minimize these compounding errors; less uncertainty is introduced due to the elimination of one of the differentiations of the measured star counts.  The column density is related to the volume mass density as follows.
\begin{equation} \label{eq:GD4-39amod}
\Sigma(z)=\int_{-z}^{z}\rho(z')dz'
\end{equation}
By using the symmetry of the vertical distribution, the column density within a distance z from the Galactic plane is obtained:
\begin{equation} \label{eq:GD4-39bmod}
\Sigma(z)=-\frac{1}{2\pi G\nu}\frac{\partial(\nu\overline{v^2_z})}{\partial z}
\end{equation}

	In order to reduce the uncertainty in the determination of the local Galactic disk mass density, other techniques have been employed which avoid the compounded differentiation associated with the method previously discussed; each approach possesses separate benefits and uncertainties.  Bahcall (1984b, 1984c, hereafter Ba84b and Ba84c respectively) employed a technique to determine the mass density assuming an isothermal equilibrium distribution function for the separate stellar components in the local Galactic disk; this method is explained in detail in Section \ref{sec:pro2}.  A variant of the local column density determination, explored thoroughly in Section \ref{sec:con2}, was employed by Kuijken and Gilmore (1989b, hereafter KG89b) in which a variational form of the potential in the local solar neighborhood is assumed and optimized in comparison to observational data.  Several other approaches are also briefly discussed.

Another completely different approach used to explore the existence of local dark matter in the Galactic disk utilizes gravitational microlensing of stars by lens objects in the foreground.  In particular, this provides a direct method of studying dark matter acting as the foreground lens.  This will be further discussed in Section \ref{sec:micro}.

\vspace{8mm}
\section[Proponents for Dark Matter in the Galactic Disk]{Proponents for Dark Matter\\ in the Galactic Disk} \label{sec:pro}

	The original proponent for the existence of unobserved, dark matter was Oort (1932, hereafter O32).  By attaining information about the gravitational density distribution of the Galactic disk, it appeared that a significant part of the local Galactic disk mass remains unaccounted for observationally and in particular is strongly concentrated in the Galactic plane.  More recent explorations have further investigated and attempted to detect the dark matter in the local Galactic neighborhood using such methods as the comparison of measured gravitating matter to the estimated amounts of visible matter.

\subsection{Exploration Renewed} \label{sec:pro2}

	Later, Ba84b and Ba84c used a different technique to again investigate the observationally missing matter in the Galactic disk.  These studies into F dwarfs and K giants also supported Oort's previous discovery of disk dark matter, but now claimed that approximately half of the mass density in the solar neighborhood of the disk is dark.

	In order to reduce the uncertainty in the determination of the local Galactic disk mass density, Ba84b and Ba84c employed an alternate technique in determining the mass density, avoiding the compounded differentiation involved with the method previously discussed.  It is based on the claim that observationally, individual disk components are reasonably approximated by isothermal density profiles; this assumption included stars, dust, and Ba84b and Ba84c further extrapolated this to the unobserved dark matter as well.  

	This method first entails looking at separate isothermal components of the mass density and assuming that each distinct component has a mean-square vertical velocity that is independent of height, z.  This assumption enables the $\overline{v_z^{2}}$ term in equation \ref{eq:GD4-36} to be removed from the differentiation, allowing a simple expression for the number density to be obtained assuming the zero-reference of the vertical potential is the plane of the disk.
\begin{equation} \label{eq:GD4-40a}
\nu _k(z)=\nu _k(0)\ e^{-\Phi(z)/\overline{v^2_z}}
\end{equation}
Similarly, an expression can be obtained for the mass density by using the stellar distribution function for an isothermal population which has the following form.
\begin{equation} \label{eq:GA10.58}
f_{z,k}(E_z)=\frac{\rho _k(0)}{\sqrt{2\pi \overline{v^2_z}}}\ e^{-E_z/\overline{v^2_z}}
\end{equation}
This distribution leads directly to a mass density of the form:
\begin{equation} \label{eq:GA10.60}
\rho _k(z)=\rho _k(0)\ e^{-\Phi(z)/\overline{v^2_z}}
\end{equation}
The superposition of each mass density component provided by equation \ref{eq:GA10.60}, can then be utilized to obtain the total mass density and finally a formula for the potential of the mass distribution using Poisson's equation (\ref{eq:GD4-37}).  Given the stellar count information for different populations as well as the fact that both the number density and mass densities have the same form, then the observed density profiles and isothermal model density profiles can be compared for each component.  A discrepancy between the calculated and observed density profiles implies the existence of missing mass.

	Ba84b and Ba84c attempted to rectify the discrepancy by claiming the missing mass was dark matter; by adding a single isothermal dark matter component and adjusting its density in the center of the plane to accommodate the observed profile for a tracer population, an estimate for the dark matter in the Galactic disk could be obtained.  Because the dark matter is fit with an isothermal equilibrium model, the determination of the dark matter component varies with the assumption of the velocity dispersion of the dark matter; larger dispersions result in smaller central densities for dark matter.

	The results obtained for F stars and K giants resulted in an average total local mass density of $\rho_0=0.18\pm 0.02 M_\odot /pc^3$.  This dynamical estimation of the mass density is substantially larger than estimations of the visible density of matter near the sun, apparently indicating the existence of Galactic disk dark matter.  

	A couple significant issues are associated with a study such as this.  First, the assumption that various tracer populations are in isothermal equilibrium must be verified observationally by showing the velocity distributions are Maxwellian.  In particular, a certain stellar population may actually include more than one isothermal component.  Furthermore, the extended assumption that the dark matter is also in equilibrium with an isothermal profile is ad hoc since the form of dark matter is not yet known. 

	One benefit of the Ba84b and Ba84c studies is that they accommodate different stellar populations.  Because variant stellar types have different $\overline{v_z^2}$, the vertical density profile for each population is distinct.  Furthermore, a model of the density profile, such as an isothermal equilibrium distribution, must first be calibrated to observations.  By decomposing matter in the solar neighborhood into separate isothermal populations, a broader more precise understanding can be obtained.  Despite the benefits, the systematic uncertainties were significant and unavoidable.  Surveys were proposed and later obtained in order to minimize these systematic errors.  

	The Bahcall, Flynn, and Gould study (1992, hereafter BFG92) resulted from these proposals in which detailed and accurate data were obtained for a K giant sample in the South Galactic Pole (SGP).  By investigating this tracer population of 125 bright K giants, it was concluded that the amount of local Galactic disk dark matter was statistically significant, but not conclusive evidence for disk dark matter.  Bahcall and his colleagues claimed that a model excluding a dark matter disk component would be inconsistent with the results of their study to within about $1.5\sigma$, an 86\% confidence level; therefore, the models which included no contingency for any missing matter only had a 14\% probability of accurately accommodating the observed data.  Although this result is substantial support of the claim for evidence in the local Galactic disk, it must not be taken as a definitive result or irrefutable evidence for the existence of dark matter in the disk; even BFG92 makes the caveat that this was a single study of an individual population and skepticism is justifiable.

	This study emphasized the high probability against a model without a disk dark matter component and further investigated various models for the dark matter such as Cold Dark Matter (CDM) at 4 km/s, Hot Dark Matter (HDM) at 20.3 km/s, and also models in which the dark matter mimics the visible matter distribution; the latter best-fit model indicated that over 50\% of the disk by mass is dark matter.  In comparison to Ba84b and Ba84c, the BFG92 study had a similar general method with a couple refinements to the theory, however this latter study resulted in a somewhat larger amount of dark matter with weaker confidence.

	The key attributes to this study were the care in minimizing uncertainties, biasing, and systematic effects, while maximizing signal-to-noise.  To avoid selection bias in the sample determination, the sample of 125 K giants were chosen prior to the theoretical analysis and used stringent selection criteria.  In addition, the systematic and random uncertainties in the data were thoroughly addressed.  Furthermore, the BFG92 analysis of the previous results by other groups such as Kuijken and colleagues seemed to support evidence for the widely varying conclusions of these separate studies; this is discussed in detail in Section \ref{sec:disc}. 
 
	An additional effect that must be considered is that the observational density determinations of visible matter do have significant error, especially the ISM component.  The most extreme ranges for possible ISM components were investigated by BFG92 to determine the effect on the previously mentioned results.  It was determined that the models that included no contingency for missing matter had a 20\% probability of accurately accommodating the observed data, as compared to the previous value of 14\%.  Thus, the support for dark matter is still significant despite possible uncertainties such as those in the amount of ISM contributing to the visible mass of the disk.

	In order to properly verify this claim of dark matter in the disk, BFG92 discussed several avenues possible to reduce the uncertainties and improve the calculation of the total amount of matter in the local Galactic neighborhood.  First, increasing the sampled stellar population by 75\% would aid exploring the existence of dark matter to within 2 $\sigma$ uncertainty.  Also, observing to greater altitudes, no greater than 1 kpc, would reduce uncertainties due to the increased sensitivity to the exponential density profile; higher observations would no longer concentrate on the disk properties.  Finally, metallicity information should also be obtained to avoid contamination of very old high-velocity stars in the new high altitude survey.

	As stated explicitly by BFG92, the Hipparcos mission would help enlighten this subject matter further by reducing uncertainties in the K giant data used in the experiment, enabling a new level of exploration into this subject matter.  This mission actually resulted in a few studies which disagreed with the results of Bahcall, as will be discussed in Section \ref{sec:hip}.

\subsection{Microlensing, a New Frontier} \label{sec:micro}

	An alternative approach used to explore the existence of local dark matter in the Galactic disk utilizes gravitational microlensing.  For example, some studies involve microlensing of stars in the Galactic bulge, observed as a brief, minor magnification on the order of 0.3 mag, by objects in the foreground acting as the lens.  In particular, the foreground lens can be dark matter thus providing a direct method of studying this observationally missing mass.  Interestingly, the duration of the microlensing event is proportional to $\sqrt{M_{lens}}$ and ranges from about a week to a month for masses from 0.1 to 1.0 $M_\odot$.  

	Without any dark matter in the disk, there is a modest rate of such microlensing events by lower mass objects, on the order of 4 per year per $10^6$ bulge stars.  However, if there were significant amounts of brown dwarfs, Jupiter-sized objects, and low mass dark matter in the disk, Paczynski (1991, hereafter Pa91) states this would increase the rate significantly to on the order of 17 per year per $10^6$ bulge stars; an observable and measurable effect.  The main issue with a project such as this is the enormous amount of time required to monitor populations continuously, often on the order of a few months because of the low yearly rates expected. 

	As an extension of the previous investigations, Gould, Bahcall, Flynn and others became involved in microlensing investigations.  One of these studies, Gould et. al. (1996, hereafter GBF96), explored a sample of 257 Galactic disk dwarfs using HST WFPC2 and PC1 images.  One main conclusion from this study was that the extrapolated lensing rate could not be explained by the visible matter; however, one cannot determine if the increased rate is a result of lensing dark matter in the disk or halo.  An additional conclusion to the microlensing investigation was that estimates of the dispersion indicate the M dwarfs in the sample are either anomalous and move slower than stars in the thick disk, or there is a substantial amount of unobserved dark matter in our Galactic disk.  This approach using microlensing to explore the existence of dark matter only touches on this aspect of probing for missing matter.

\vspace{8mm}
\section[Opponents of Dark Matter in the Galactic Disk]{Opponents of Dark Matter\\ in the Galactic Disk} \label{sec:con}

	Despite the early advances supporting the existence of substantial amounts of dark matter in the Galactic disk, other investigations obtained opposite conclusions.

\subsection{The Debate} \label{sec:con2}

	One of the first groups to contradict the claim of significant dark matter in the Galactic disk was Bienaym\'e at. al. (1987, hereafter BRC87).  By constructing a model of the Galactic density distribution, the theoretical rotation curve was ascertained and compared with observations.  The mass-density model involved four different components: ISM disk with both a radial and vertical exponential profile, a point mass approximated bulge, a dark matter halo necessary to accommodate the flat rotation curve at large radii, and a dark matter disk component also with a double exponential in the radial and vertical directions.  Therefore, by optimizing the disk dark matter component to fit the rotation curve observations, an estimation of the dark matter in the disk could be obtained.  

	BRC87 found that the disk stellar population model resulted in a potential that reproduced the rotation curve without requiring any dark matter in the Galactic disk.  The dynamically determined mass density was determined to be $0.105\pm 0.015 M_\odot/pc^3$ which agrees with the estimated observed local mass, $0.10 M_\odot/pc^3$.  Therefore little to no significant dark matter is necessary in the local Galactic disk, a conclusion which conflicts with the substantial amount of disk dark matter suggested by Ba84b and Ba84c, as well as the later conclusions by BFG92.  In particular, the best fit model required only $0.01 M_\odot/pc^3$ of dark matter with a scale height of 600 pc; a slight augmentation of the dark halo component of the Galactic dark matter would accommodate this insignificant disk matter discrepancy.  

	Additional opponents to the existence of Galactic disk dark matter, KG89b utilized the variational principle in the determination of the local column density. Their dynamical investigations into the surface mass density in the solar neighborhood indicate that there is no significant dark matter in the disk of the Galaxy. 

	To explore the method of this investigation, first assume the radial and vertical motions are independent and temporarily neglect the $v_R-v_z$ coupling; also assume the system is in a steady state.  Using Jean's Theorem, the stellar distribution, f(z,$v_z$)=f($E_z$), is expressed in terms of $E_z$, the energy involved in the z motion. 
\begin{equation}  \label{eq:energyz}
E_z=\frac{1}{2}v_z^2+\Phi_z(z)
\end{equation}
The stellar number density can then be ascertained
\begin{equation} \label{eq:GA10.49a}
\nu(z)=\int_{-\infty}^{\infty} dv_zf(z,v_z)=2\int_{0}^{\infty} dv_zf(z,v_z)
\end{equation}
By changing variables to $E_z$ given by equation \ref{eq:energyz}, the stellar density becomes 
\begin{equation} \label{eq:GA10.49b}
\nu(z)=2\int_{\Phi_z(z)}^{\infty} dE_z \frac{f_z(E_z)}{\sqrt{2(E_z-\Phi_z(z))}}
\end{equation}
By recognizing this has the form of the Abel integral equation, as discussed in Appendix Section \ref{sec:abel}, we can invert this to obtain an equation for the distribution function.
\begin{equation} \label{eq:GA10.50}
f_z(E_z)=-\frac{1}{\pi} \int_{E_z}^{\infty} \frac{d\nu}{d\Phi_z} \frac{d\Phi_z}{\sqrt{2(\Phi_z-E_z)}}
\end{equation}

	Next, a variational form of the potential must be assumed and optimized using Equation \ref{eq:GA10.50} to obtain the best fit to the observed distribution.  Once optimized, the final potential can then be utilized in Poisson's equation (\ref{eq:GD4-37}) to calculate the mass density, $\rho$. In order to determine the variational form of the potential, the gravitational force arising from the separate components of the disk and halo must be ascertained.  After superimposing them to form the net gravitational force, this can then be used to determine a general form of the potential. 
	
	For the disk component with a $z_0$ disk scale height, the density profile was assumed to have the form, $\rho = \rho_0 e^{-|z|/z_0}$, which can be utilized along with Gauss' Law to determine a functional form of the gravitational force from the disk.
\begin{equation}
F_D=8\pi G\rho_0 z_0 (e^{-z/z_0}-1)
\end{equation}
For the limit $z\ll z_0$, $F_D\simeq -8\pi G\rho_0 z_0 (\frac{z}{z_0})=C_D(\frac{z}{z_0})$, while for the limit $z\gg z_0$, $F_D\simeq -8\pi G\rho_0 z_0=C_D$; the constant $C_D\equiv -8\pi G\rho_0 z_0$ is essentially the disk surface density, while $C_D/z_0$ is basically the equatorial volume density since it is proportional to $\rho_0$.  A functional form of the force that supports these asymptotic limits for the disk distribution has the form:
\begin{equation}
F_D=\frac{C_Dz}{\sqrt{z^2+z_0^2}}
\end{equation}

In order to determine the form of the force due to the halo component, assume the equipotential surfaces are nearly spheroidal where $a\approx 1$.  Thus $\Phi_H(R,z)\equiv\Phi_H(R^2+z^2/a^2)$.  Then letting $r\equiv R^2+z^2/a^2$ the chain rule can be employed.
\begin{equation}
F_H=-\frac{\partial\Phi_H}{\partial z}=-\frac{\partial\Phi_H}{\partial r}\frac{\partial r}{\partial z}=\frac{\partial\Phi_H}{\partial r}\frac{2z}{a^2}
\end{equation}
The factor $C_H$, which is essentially the spheroidal halo component mass density, can now be introduced for simplification.
\begin{equation}
C_H=\frac{1}{a^2}\Big[\frac{\partial\Phi_H}{\partial r}\Big]_{R_0}
\end{equation}
\begin{equation}
F_H=2C_Hz
\end{equation}

The total force is then the superposition of the final expressions of the gravitational force from both the disk and halo components. 
\begin{equation}
F_z=\frac{C_Dz}{\sqrt{z^2+z_0^2}} + 2C_Hz
\end{equation}
Therefore, the potential has the following general form
\begin{equation}
\Phi_z(z)=C_D(\sqrt{z^2+z_0^2}-z_0)+C_Hz^2
\end{equation}
where $z_0$ is the disk's vertical scale height and $C_D$ versus $C_H$ determine the relative proportion of the disk and halo density components.  This equation for the potential typifies a main controversy of the study of dark matter in the local Galactic disk.  The question not only includes whether there is dark matter locally, but whether it is related to the disk or spheroidal halo components of the Galaxy.  

	By substituting $E_z$ and the assumed form of $\Phi_z(z)$ into equation \ref{eq:GA10.50} for the $f(E_z)$,  the distribution function can be determined in terms of the parameters defining the potential.  The optimum parameters that result in agreement of the observed and modeled stellar distributions then provide the final potential of the system.  Once the potential is known, then the mass density and column density can easily be determined.

	As an additional measure, KG89a and KG89b made a correction for the $v_R-v_z$ coupling that was neglected.  By not omitting this term, referred to as the tilting term by KG89a, it can be combined with the previously calculated one dimensional force, resulting in an overall effective force.  The correction is negative, depending on the shape and the change in orientation of the velocity ellipsoid in space; it can only be roughly estimated due to the unknown dependence of the orientation with vertical height, z.  KG89b claims that neglecting this contribution can lead to underestimating the column density by as much as 20\%.	

	The goal of a study such as this is to separate the disk and halo components so that they can be compared to observed estimations therefore enabling a specific determination of the amount of disk dark matter; thus, the accurate estimation of the optimum values for $C_D$, related to the surface density of the disk, and $C_H$, related to the mass density of the spheroidal halo, is of main importance.  KG89b, using information about the rotation curve of the Galaxy, applied a constraint inversely relating these constants, $C_H = 0.041 - 0.0094C_D \pm 0.008$ given $z_0 \sim 180 pc$,  and found the resulting 1 $\sigma$ range for the total column density to be $46\pm 9 M_\odot /pc^2$.  However, Gould (1990, hereafter Go90), a collaborator of Bahcall, found issues with the study.   Using the same K dwarf data and assumptions from the KG89b study, Gould found the total column density to be approximately 1 $\sigma$ greater, $54\pm 8 M_\odot /pc^2$; the main criticism was that the KG89b result was not obtained using the robust maximum likelihood method. 
	
	Kuijken and Gilmore (1991, hereafter KG91) revisited the earlier analysis with a study of a single trace population of 512 K dwarfs and estimated the disk mass component to be $48\pm 9 M_\odot /pc^2$ which appears to agree well with the observed mass density estimation, $48\pm 8 M_\odot /pc^2$ consistent with no missing disk matter.  Kuijken and his colleagues concluded that there was no significant dark matter component in the disk since the observed and calculated disk mass density determinations were in considerable agreement.  Kuijken (1991, hereafter K91), using data obtained from a 512 faint K dwarf stellar population at the SGP, further investigated this topic by modeling dynamical stellar motions and found there was no mass discrepancy in the observed disk matter within 160 pc of the Galactic Plane.  

	In contrast to the assumption made by Bahcall and his colleagues that the distribution function for each tracer population component was isothermal, a benefit of this method is that it avoids applying this isothermality condition on the stellar population under investigation.  However, an obvious limitation is that these studies cannot predict the density of all types of stars in the solar neighborhood of the disk since they include only a single tracer population.   

	In contrast to his previous work in 1992 with Bahcall and Gould as proponents to the Galactic disk dark matter, Flynn became involved in a study with Fuchs (1994, hereafter FF94) which extended the BFG92 investigation to local K Giants.  The result of this extended study was that the observational data could be accommodated within 1 $\sigma$ uncertainty by the known disk matter in agreement with results previously obtained by KG89b and Kuijken and Gilmore (1989c, hereafter KG89c).  FF94 does mention that the fit can be improved by adding a small component of disk dark matter; if distributed proportionally to the visible matter then the local column density is $52\pm 13 M_\odot /pc^2$ as opposed to $49\pm 9 M_\odot /pc^2$ visible.  Interestingly, Flynn was a part of the original team, BFG92, to support the existence of significant dark matter, but is now substantiating conclusions that greatly reduce the significance of dark matter in the disk and even validate the claims that a dark matter component to the Galactic disk is not necessary to accommodate the observations.  Another more recent article that will be discussed in the next section, Holmberg and Flynn (2000, hereafter HF00), also includes Flynn and further disputes the existence of Galactic disk dark matter.

\pagebreak
\subsection{Discoveries by Hipparcos} \label{sec:hip}

	The introduction of ESA's Hipparcos satellite has brought much enlightenment to the debate involving studies by scientists such as Pham (1997, hereafter Ph97), Cr\'ez\'e et. al. (1998, hereafter CCBP98), and HF00.  Among other things, this satellite has been able to attain accurate distance and proper motions for a complete set of nearby stars as well as obtain an improved Luminosity Function from which the local visible matter can be better estimated.  This resulting data allows investigations of much higher accuracy, improving the comparison between the amount of observed matter to the calculations of the amount of gravitating matter; the discrepancy being the dark matter.  Each of these studies based on Hipparcos data supports the view that there is no compelling evidence in support of the existence of significant dark matter in the Galactic disk.

	Using data from the Hipparcos catalogue, Ph97 studied the F stellar tracer population and measured the density using a slightly different technique employing the isothermal distribution assumption.  After neglecting the $v_R-v_z$ coupling, calculations of the disk scale height and velocity dispersions from the Hipparcos data are then utilized along with the assumption of isothermality to determine the local disk mass density:
\begin{equation}
\rho(0)=\frac{\sigma_z^2}{2\pi Gh_z^2}
\end{equation}
The scale height was estimated to be $h_z=165\pm 7 pc.$  The local volume mass density calculated using this method, $0.11\pm 0.01 M_\odot /pc^3$, is in agreement with the observed mass density $0.10 M_\odot /pc^3$; no evidence indicates the existence of significant dark matter in the local Galactic disk.
	
	Conducting a study using the results from the Hipparcos satellite data on A and F stars, CCBP98 utilized the method of vertical kinematics to determine the local density of matter; these tracer populations provided a vertical density change significant enough for the local density to be measured.  In particular, they obtain a particularly low estimation for the total local volume density, $0.076\pm 0.015 M_\odot /pc^3$, possibly due to a slightly different determination from local data within a sphere of on the order of 100 pc radius.  One benefit suggested is that many of the other studies have attempted to find the upper density bounds while this gives a lower limit for the total density.  Their results support the claim that no dark matter exists in the local disk.

	HF00 also supports the claim against the existence of significant amounts of dark matter in the Galactic disk.  Using parallaxes and proper motions of A and F stars from the Hipparcos catalogue, the velocity distribution of the tracer population along with a model of the local disk potential can be used to determine the vertical density decline which can then be compared to the observed data to ascertain the applicability of the model.  HF00 estimated the local dynamical mass density of the Galactic disk to be $0.098\pm 0.011 M_\odot /pc^3$ versus the estimated observed density of $0.095 M_\odot /pc^3$; therefore because the kinematic calculation of the mass density and the observed density were comparable, there was no convincing evidence for substantial amounts of dark matter in the Galactic disk.

	Another method for ascertaining the shape of dark matter in the Galaxy was employed using the Hipparcos proper motion data.  By determining the velocity distribution of local stellar populations and estimating the density gradients and finally the shape of the potential, Bienaym\'e (1999) was able to deduce that the Galactic potential was not significantly flat and that therefore the dark matter component is not confined to the Galactic disk.  As is apparent, the more recent evidence on this topic, supported by the Hipparcos data, indicates that there is no significant dark matter in the Galactic disk.

\vspace{8mm}
\section{Discussion} \label{sec:disc}

	As previously discussed, there have been varied approaches to the investigation into the existence of dark matter in the Galactic disk resulting in discordant conclusions both for and against this claim.  The methods, assumptions, and conclusions of several groups of scientists have been explored and now will be addressed as a group to determine the conclusion that best accommodates all of the issues.

	Following the claim by Ba84b and Ba84c that a majority of the disk mass consists of dark matter, KG89c and Cr\'ez\'e et. al. (1989, hereafter CRB89) found several issues to criticize.  In particular, KG89c claimed that the K giant population used by Bahcall involved a superposition of two components; half CDM at 11km/s and half HDM at 20km/s; comparisons with the data from Oort did not take this properly into account leading to overestimation of the disk mass and therefore inflated dark matter estimations.  BFG92 verified this problem, but claimed the effect on their results was not as significant as previously believed and does not come into effect at all in the updated study BFG92.

	In addition, CRB89, the second part of the study by BRC87, addresses the discrepancies between those results and the opposite claims of Ba84b and Ba84c asserting the existence of significant amounts of dark matter.  The main criticism made by CRB89 was regarding the assumption by Ba84c that the stellar density measurements obtained from smoothed magnitude counts obey Poisson statistics; concerns were raised regarding bin correlations.  CRB89 claims that because of these errors, the Bahcall models were biased towards more missing mass, and are actually compatible with observed data implying no need for significant Galactic disk dark matter.  In fact, they make the claim that any of those models considered, even a model with no missing mass, is compatible with the data to within $2\sigma$.  In order to address this issue, BFG92 used explicit stellar counts instead, therefore avoiding this problem; although the results obtained contained more uncertainty, BFG92 still obtained results in general agreement with the previous conclusions of Ba84b and Ba84c indicating the need for a substantial Galactic disk dark matter component.  

	BFG92 also discusses issues that arise in work by KG89c and KG91, criticizing correction terms, physical assumptions, and statistical methods applied in their studies.  In particular, BFG92 states that without these ad hoc corrections, the results from Kuijken et. al. studies would have also supported the existence of dark matter in the local galactic disk.  Furthermore, BFG92 claims that significant systematic errors in the work by BRC87 have resulted in the conclusion of no disk dark matter.

	It is apparent that each method employed to study the possible existence of Galactic disk dark matter does have associated with it many issues relating to the assumptions made in the modeling of the disk density as well as the estimations of the final uncertainties relating to the data.  Each side has criticisms relating to the different aspects of other studies probing this subject.  The answer to the question of the existence of dark matter will therefore not be inherently proven or disproven by one such study, but the numerous studies must be analyzed as a whole to decipher the evidence in support of the overall claims.  

	As a review of the results of the majority of studies discussed, Tables \ref{columndensity} and \ref{volumedensity} are provided, listing the surface and volume mass densities respectively.\footnote{Extreme variations were obtained by Ba84b and BFG92 due to the diverse models investigated.  The values in Table \ref{columndensity} are taken from the quoted best fit for dark matter proportional to the visible matter.} 

\begin{table}[t]
\caption{Summary of column density results from various studies of the local solar neighborhood. } 
\vspace{3mm}
\begin{center}
\begin{tabular}{|l|c|c|c|} 
\hline
& $\Sigma_{Obs}$ & $\Sigma_{Total}$ & \\
Study & $M_\odot /pc^2$ & $M_\odot /pc^2$ & Dark Matter\\
\hline
\hline
O32 & - & $51.5^{\ast}-80.4$ & Yes \\
Ba84b & - & $67\pm 5$ & Yes \\
KG89c & $48\pm 8 $ & $46\pm 9$ & None \\
Go90 & $48\pm 8$ & $54\pm 8$ & $\ddagger$ \\
KG91 & $48\pm 8$ & $48\pm 9$ & None \\
BFG92 & - & $83.9 $  & 53\%\\
FF94  & $49\pm 9$ & $52\pm 13$ & 0-20\%\\
\hline
\multicolumn{4}{|l|}{ \scriptsize $\ast$ Truncated at 200 pc.}\\
\multicolumn{4}{|l|}{ \scriptsize $\ddagger$ No conclusion regarding significance of disk dark matter.}\\
\hline
\end{tabular} 
\label{columndensity}
\end{center}
\end{table}
\vspace{5mm}

\begin{table}[t]
\caption{Summary of volume density results from various studies of the local solar neighborhood.}
\vspace{3mm}
\begin{center}
\begin{tabular}{|l|c|c|c|} 
\hline
& $\rho_{Obs}$ & $\rho_{Total}$ & \\
Study & $M_\odot /pc^3$ & $M_\odot /pc^3$ & Dark Matter\\
\hline
\hline
O32 & $\ddagger$ & $0.092\pm 0.018$ & Yes \\
Oort $\dagger$& $0.114-0.140 \dagger$ & $0.15 \dagger$ & Yes \\
Ba84b & $0.114-0.140 \dagger$ & $0.18\pm 0.02$ & 48\% \\
BRC87 & 0.10 & $0.105\pm 0.015 $ & $0-0.03, 0.01 M_\odot/pc^3$ best\\
Ph97 & 0.10 & $0.11\pm 0.01$ & None \\
CCBP98 & $0.085\pm 0.02$ & $0.076\pm 0.015$ & None \\
HF00 & 0.095 & $0.098\pm 0.011$ & None \\
\hline
\multicolumn{4}{|l|}{ \scriptsize $\ddagger$ $0.038 M_\odot /pc^3$ stellar to 13.5 mag, $0.05 M_\odot /pc^3$ nebulous, etc.}\\
\multicolumn{4}{|l|}{ \scriptsize $\dagger$ Obtained from BT94. }\\
\hline
\end{tabular} 
\label{volumedensity}
\end{center}
\end{table}
\pagebreak
\vspace{8mm}
\section{Conclusion} \label{sec:conc}
	The investigation into the amount and significance of dark matter in the solar neighborhood of the Galactic disk has been a sustained and dramatic debate, especially the past two decades.  Due to the extensiveness of the studies devoted to exploring the possible existence of local disk dark matter, the goal of this investigation was to gain a broad understanding of the initial debates in this subject matter as well as a general insight into some more recent advances.  

	Previously, the proponents for Galactic disk dark matter had a rise in support by the studies of Bahcall and his colleagues, but have not sustained continued support for this view.  In fact, it appears that Flynn, previously a strong proponent of the Galactic disk dark matter with BFG92, has greatly altered his view in FF94 by stating that this study requires little to no dark matter in agreement with previous Kuijken et. al. results.  Later in HF00, Flynn further abandoned the original view for dark matter by claiming there is no convincing evidence for substantial amounts of dark matter in the Galactic disk.

	It appears the opponents against the existence of dark matter in the Galactic disk are gaining more ground in support of their position, especially since the Hipparcos data became available; several studies have provided support to the claim that significant dark matter in the disk is not required to accommodate observations.

\pagebreak
\section{Appendix: Jeans Equation Simplification} \label{sec:app}
	In the Introduction, Section \ref{sec:intro}, the mass density $\rho$ in the Galactic disk is provided by equation \ref{eq:GD4-38mod}.  A main component of this derivation is the demonstration that the first two terms in the cylindrical form of Jeans equation (\ref{eq:GD4-29cmod}) are insignificant in the Galactic disk system.  The first term is unimportant due to the assumption that the system is in steady state, while the negligibility of the second term $\frac{1}{R}\frac{\partial}{\partial R}(R\nu\sigma^2_{Rz})$ is what will now be shown.  This particular problem and its initial assumptions are posed as problem 4.5 in BT94.

	In the Galactic disk near the sun where $\frac{|z|}{R}\ll 1$, assume that the principle axes of the velocity ellipsoid are parallel to the spherical coordinate unit vectors. To begin, the cylindrical coordinates (R,$\phi$,z) and spherical coordinates (r,$\phi,\theta$) have the following correspondence:
\begin{eqnarray}
\phi & = & \phi \\
R & = & rsin\theta \\
z & = & rcos\theta
\end{eqnarray}
By taking the first derivative of the equations for R and z, the radial and vertical velocities can be obtained as a function of the spherical coordinates.
\begin{eqnarray}
v_R & = & R' = rcos\theta\theta '+r'sin\theta  \\
v_z & = & z' = -rsin\theta\theta '+r'cos\theta
\end{eqnarray}
Given that $v_r=r'$ and $v_{\theta}=r\theta '$, these equations for the velocities in the R and z directions can be further transformed:
\begin{eqnarray}
v_R & = & v_{\theta}cos\theta +v_r sin\theta  \\
v_z & = & -v_{\theta}sin\theta +v_r cos\theta
\end{eqnarray}
The product of the radial and vertical velocities in cylindrical coordinates then becomes:
\begin{equation}
v_Rv_z = sin\theta cos\theta(v_r^2-v_{\theta}^2)+v_rv_{\theta}(cos^2\theta-sin^2\theta)
\end{equation}

After taking the mean and applying $\overline{v_rv_{\theta}}=0$ due to the motions being independent, then the mixed radial and vertical velocity dispersion can be determined.
\begin{equation}
\overline{v_Rv_z} = \overline{sin\theta cos\theta}(\overline{v_r^2}-\overline{v_{\theta}^2})
\end{equation}
In the region near the Galactic plane where $\frac{|z|}{R}\ll 1$, it follows that $\theta\simeq\frac{\pi}{2}$ so that we can take advantage of the relation $sin\theta\simeq 1/sin\theta$ to provide an expression for the angular dependence. 
\begin{equation}
sin\theta cos\theta \simeq \frac{cos\theta}{sin\theta} = \frac{z}{R}
\end{equation}
Therefore the coupled velocity dispersion simplifies.
\begin{equation}
\sigma^2_{Rz} = \overline{v_Rv_z} \simeq \overline{\frac{z}{R}}(\overline{v_r^2}-\overline{v_{\theta}^2})
\end{equation}

	Now utilizing this expression for the coupled velocity dispersion, it can be shown that the term, $\frac{1}{R}\frac{\partial}{\partial R}(R\nu\sigma^2_{Rz})$, is negligible in the cylindrical Jean's equation (\ref{eq:GD4-29cmod}).  First, explore the order of magnitude estimations of the term of interest and another of the remaining terms in equation \ref{eq:GD4-29cmod}.
\begin{eqnarray} 
\label{eq:term2}
Eq. \ref{eq:GD4-29cmod}, 2^{nd} term   &  \frac{1}{R}\frac{\partial}{\partial R}(R\nu\sigma_{Rz}^2) \simeq \frac{z}{R}\frac{\partial}{\partial R}[\nu(\overline{v_r^2}-\overline{v_{\theta}^2})] \sim \frac{z\nu(\overline{v_r^2}-\overline{v_{\theta}^2})}{RR_d}\\
\label{eq:term3}
Eq. \ref{eq:GD4-29cmod}, 3^{rd} term   &  \frac{\partial(\nu\overline{v^2_z})}{\partial z} \sim \frac{\nu\overline{v^2_z}}{z}
\end{eqnarray}
In comparison, the term in equation \ref{eq:term2} is much smaller by a factor of magnitude $\frac{z^2}{RR_d}$ than the term in equation \ref{eq:term3}.  Because we are in the regime near the Galactic plane where $|z|/R\ll 1$, it follows that the term in equation \ref{eq:term2}, which is the second term in the cylindrical Jeans equation \ref{eq:GD4-29cmod}, is negligible and therefore also assuming steady state equilibrium, this Jean's equation may be simplified to the one-dimensional form in equation \ref{eq:GD4-36}.

\section{Appendix: Abel's Equation} \label{sec:abel}

	Section \ref{sec:con2} discusses the method employed by KG91 in determining the local column density in the Galactic disk; in particular, the stellar number density, provided as equation \ref{eq:GA10.49b}, has the form of the Abel integral equation.  BT94 reviews the format of the Abel integral equations in detail.  Given an integral equation of the form 
\begin{equation} \label{eq:GD1B-59a}
F(x)=\int_x^\infty \frac{G(t)dt}{(t-x)^{\alpha}}
\end{equation}
then an expression for G(t) follows directly from the Abel integration relation.
\begin{equation} \label{eq:GD1B-59b}
G(t)=-\frac{sin\pi\alpha}{\pi} \bigg[\int_t^\infty \frac{dF}{dx} \frac{dx}{(x-t)^{1-\alpha}} - \lim_{x \rightarrow \infty} \frac{F(x)}{(x-t)^{1-\alpha}}\bigg]
\end{equation}

Equation \ref{eq:GD1B-59b}  can immediately be applied to equation \ref{eq:GA10.49b} for the number density by recognizing the following correlations $x \rightarrow \Phi_z$, $t \rightarrow E_z$, $F \rightarrow \nu$, $G \rightarrow \sqrt{2}f_z$, and $\alpha \rightarrow \frac{1}{2}$.  Using the boundary condition specifying that the density falls off as vertical distance extends to infinity, then one can immediately ascertain the following relation for $f_z(E_z)$:

\begin{equation} \label{eq:GA10.50app}
f_z(E_z)=-\frac{1}{\pi} \int_{E_z}^{\infty} \frac{d\nu}{d\Phi_z} \frac{d\Phi_z}{\sqrt{2(\Phi_z-E_z)}}
\end{equation}

\vspace{20mm}
\pagebreak

\end{document}